\documentclass[a4paper]{article}
\usepackage{amsmath,amssymb}
\title{The Indefinite Self-Dual Metrics and Painleve Equations}
\author{Shoji Okumura (okumura@math.sci.osaka-u.ac.jp)\\Math. Sci. Osaka Univ., Japan}
\date{}

\begin{document}
\maketitle

\newtheorem{thm}{Theorem}[section]
\newtheorem{prop}[thm]{Proposition}
\newtheorem{rem}[thm]{Remark}
\newtheorem{definition}[thm]{Definition}
\newtheorem{cor}[thm]{Corollary}
\newtheorem{lemma}[thm]{Lemma}

\begin{abstract}
We classify the $SU(2)$-invariant anti-self-dual metrics with a signature 
$(+,+,-,-)$. The metrics are specified by a solution of Painlev\'e VI, V, 
III or II. Moreover we show the geometric meaning of the metrics 
specified by each type of Painlev\'e functions. 
\end{abstract}

\section{Introduction}

The aim of this paper is to classify the anti-self-dual metrics in real 
dimension four admitting an isometric action of $SU(2)$ with generically 
three-dimensional orbits. In this paper, we study not only the definite 
metrics, but also the indefinite metrics with a signature $(+,+,-,-)$. 

Hitchin \cite{Hitchin} shows that the $SU(2)$-invariant anti-self-dual 
metric is generically specified by a solution of Painlev\'e VI with two 
complex parameters. He used the twistor correspondence \cite{Penrose} to associate the 
anti-self-dual equation and Painlev\'e equation. On twistor space, the 
lifted action of $SU(2)$ determines a pre-homogenious action of $SU(2)$, 
and it determines a isomonodromic family of connections on 
$\mathbb{CP}^1$, and then we have Painlev\'e equations. In this framework, 
Dancer \cite{Dancer} shows that the diagonal scalar-flat K\"ahler metric 
is specified by a solution of Painlev\'e III with parameters $(4,4,4-4)$. 
The author \cite{oku}  shows that the $SU(2)$-invariant anti-self-dual 
hermitian metric is specified by a solution of Painlev\'e III with one 
complex parameter. 

If the metric is definite, then the anti-self-dual 
equation reduce to either Painlev\'e VI or III. On the other hand, we 
show that, if the metric has a a signature $(+,+,-,-)$, then the 
anti-self-dual equation reduce to not only Painlev\'e VI or III but also 
Painlev\'e V or II. The difference of types of Painlev\'e is due to the 
difference of reality on the twistor space. 

Painlev\'e VI is shown to be deformation equations for a linear 
problem 
\[
 \left(\frac{d}{dz}-B_1\right)\left(
  \begin{array}{c}
    y_1   \\
    y_2   \\
  \end{array}
\right)=0,
\]
where $B_1$ has four simple poles on $\mathbb{CP}^1$ \cite{JMU-I}. 
And Painleve II, III, IV, V are degenerated from Painlev\'e VI:
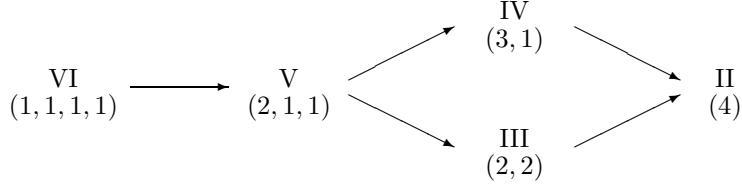
\begin{center}
\setlength{\unitlength}{1mm}
\begin{picture}(88,22)
\put(0,10){\makebox(0,0){\shortstack{VI\\$(1,1,1,1)$}}}
\put(9,11){\vector(1,0){13}}
\put(30,10){\makebox(0,0){\shortstack{V\\$(2,1,1)$}}}
\put(38,12){\vector(2,1){14}}
\put(38,10){\vector(2,-1){14}}
\put(60,19){\makebox(0,0){\shortstack{IV\\$(3,1)$}}}
\put(60,2){\makebox(0,0){\shortstack{III\\$(2,2)$}}}
\put(68,3){\vector(2,1){14}}
\put(68,19){\vector(2,-1){14}}
\put(88,10){\makebox(0,0){\shortstack{II\\$(4)$}}}
\end{picture}
\end{center}
This is the confluence diagram of poles of $B_1$, where the Roman 
numerals represent the types of Painlev\'e, and the parenthesized 
numbers represent the orders of poles of $B_1$. For example, Painlev\'e 
V is shown to be deformation equations for a linear problem with one 
double and two simple poles. Due to the reality of twistor space, the 
poles of $B_1$ makes two conjugate pairs. Therefore, the configuration 
of poles never becomes Painlev\'e IV type. 

The multiple pole of $B_1$ determines a hermitian structure or a 
structure of null surfaces, and this is the the geometric meaning of the 
metrics specified by each type of Painlev\'e functions. 

\textbf{Acknowledgements. }
The author would express his sincere gratitude to Professor Yousuke Ohyama, who
intoroduced him to the subject, for enlightening discussions.

\section{The Diagonal Anti-self-dual Equations}
In this section, we review the anti-self-dual equations on the 
$SU(2)$-invariant diagonal metrics. 

The $SU(2)$-invariant diagonal metric is written in the following form: 
\begin{equation}
    g = w_1 w_2 w_3 \, d t^2 + \frac{w_2 w_3}{w_1} \sigma_1^2 +
       \frac{w_3 w_1}{w_2} \sigma_2^2 + \frac{w_1 w_2}{w_3} \sigma_3^2 .
\end{equation}
$w_1$, $w_2$ and $w_3$ are functions of $t$, and $\sigma_1$, $\sigma_2$,
$\sigma_3$ are left invariant one-forms on each $SU(2)$-orbit satisfying
\begin{align}
  d \sigma_1 &= \sigma_2 \wedge \sigma_3 , &
  d \sigma_2 &= \sigma_3 \wedge \sigma_1 , &
  d \sigma_3 &= \sigma_1 \wedge \sigma_2 .
\end{align}
Tod \cite{Tod} showed that the anti-self-dual equations on the 
$SU(2)$-invariant diagonal metric are given by the following system: 
\begin{equation}
 \begin{split}
    \dot{ w_1} &= - w_2 w_3 + w_1 \left(\alpha_2 + \alpha_3 \right), \\
    \dot{ w_2} &= - w_3 w_1 + w_2 \left( \alpha_3 + \alpha_1 \right),  \\
    \dot{ w_3} &= - w_1 w_2 + w_3 \left(\alpha_1 + \alpha_2 \right),  \\
    \dot{ \alpha_1} &= - \alpha_2 \alpha_3 +
           \alpha_1 \left(\alpha_2 + \alpha_3 \right),  \\
    \dot{\alpha_2} &= - \alpha_3 \alpha_1 +
           \alpha_2 \left(\alpha_3 + \alpha_1 \right),   \\
    \dot{ \alpha_3} &= - \alpha_1 \alpha_2 +
           \alpha_3 \left(\alpha_1 + \alpha_2 \right),
 \end{split}   \label{d-sd}
\end{equation}
where $\alpha_1,\alpha_2,\alpha_3$ are auxiliary functions and the dots 
denote differentiation with respect to $t$. The anti-self-dual equation 
\eqref{d-sd} has a first integral
\[
 k=\frac{\alpha_1(w_2^2-w_3^2)+\alpha_2(w_3^2-w_1^2)+\alpha_3(w_1^2-w_2^2)}{
8(\alpha_1-\alpha_2)(\alpha_2-\alpha_3)(\alpha_3-\alpha_1)
}.
\]
Furthermore, if we set 
\begin{equation*}
\begin{split}
x&=\frac{\alpha_2-\alpha_1}{\alpha_2-\alpha_3},\\
q&=\frac{w_2(\alpha_1-\alpha_2)(w_2(w_1^2-w_3^2)+2\sqrt{2k}\,w_1w_3(\alpha_1-\alpha_3))}{
w_1^2(w_2^2-w_3^2)\alpha_1+w_2^2(w_3^2-w_1^2)\alpha_2+w_3^2(w_1^2-w_2^2)\alpha_3},
\end{split}
\end{equation*}
then the system \eqref{d-sd} generically reduces to a family of 
Painlev\'e VI with special parameters
\begin{equation*}
\left(\alpha,\beta,\gamma,\delta\right)=
                  \left(\frac{\bigl(\sqrt{2k} - 1\bigr)^2}{2}, k, 
                  -k, \frac{1+2k}{2} \right).
\end{equation*}
We will review the Painlev\'e equations in the appendix.

\section{The Non-diagonal Anti-self-dual Equations} \label{ODE}

We can write a $SU(2)$-invariant metric in the form
\begin{align*}
g=f(\tau) d\tau^2 + \sum _{l,m=1}^3 h_{l\,m}(\tau)\, \sigma_l \sigma_m.
\end{align*}
Using the Killing form, we can diagonalize the metric $g$ on each $SU(2)$-orbit. 
Then we can express the metric as follows:
\begin{align*}
g=(a b c)^2 dt^2+a^2d\hat{\sigma}_1^2+b^2\hat{\sigma}_2^2+c^2\hat{\sigma}_3^2,
\end{align*}
for some $t=t(\tau),$ $ a=a(t),$ $ b=b(t),$ $ c=c(t)$ and 
\begin{align*}
\left(
  \begin{array}{c}
    \hat{\sigma}_1   \\
    \hat{\sigma}_2   \\
    \hat{\sigma}_3   
  \end{array}
\right)=R(t)
\left(
  \begin{array}{c}
    \sigma_1   \\
    \sigma_2   \\
    \sigma_3   
  \end{array}
\right),
\end{align*}
where $R(t)$ is $SO(3)$-valued function.

Since $\dot{R}R^{-1}\in \mathfrak{so}(3) $,
we can write 
\begin{align*}
d\left(
  \begin{array}{c}
    \hat{\sigma}_1   \\
    \hat{\sigma}_2   \\
    \hat{\sigma}_3   
  \end{array}
\right)&=R(t)
\left(
  \begin{array}{c}
    \sigma_2\wedge\sigma_3   \\
    \sigma_3\wedge\sigma_1   \\
    \sigma_2\wedge\sigma_2   
  \end{array}
\right)+
\dot{R}\, dt\wedge\left(
  \begin{array}{c}
    \sigma_1   \\
    \sigma_2   \\
    \sigma_3   \\
  \end{array}
\right)\\
 &=\left(
  \begin{array}{c}
    \hat{\sigma}_2\wedge\hat{\sigma}_3   \\
    \hat{\sigma}_3\wedge\hat{\sigma}_1   \\
    \hat{\sigma}_1\wedge\hat{\sigma}_2   \\
  \end{array}
\right)+
\left(
  \begin{array}{rrr}
    0       &  \xi_3    & -\xi_2  \\
   -\xi_3         &  0              &  \xi_1  \\
    \xi_2         & -\xi_1    &   0     \\
  \end{array}
\right)dt\wedge
\left(
  \begin{array}{c}
    \hat{\sigma}_1   \\
    \hat{\sigma}_2   \\
    \hat{\sigma}_3   \\
  \end{array}
\right),
\end{align*}
for some $\xi_1=\xi_1(t)$, $\xi_2=\xi_2(t)$, $\xi_3=\xi_3(t)$.

If $\xi_1=0,$ $ \xi_2=0,$ $ \xi_3=0$, then the matrix $(h_{l\,m})$ can be chosen to be
diagonal for all $\tau$, and then we say that $g$ is a diagonal metric.

In the following, we mainly study the non-diagonal case.

We set $w_1=b c, w_2=c a, w_3=a b$ and 
determine $\alpha_1,\alpha_2, \alpha_3$ by
\begin{align}
\begin{split}
\dot{w}_1&=-w_2 w_3 + w_1(\alpha_2+\alpha_3),\\
\dot{w}_2&=-w_3 w_1 + w_2(\alpha_3+\alpha_1), \\
\dot{w}_3&=-w_1 w_2 + w_3(\alpha_1+\alpha_2).
\end{split}\label{sd1}
\end{align}
Then the anti-self-dual equations are as follows \cite{oku}:
\begin{align}
\begin{split}
\dot{\alpha}_1=-\alpha_2\alpha_3+\alpha_1(\alpha_2+\alpha_3)+
      \frac{1}{4}(w_2^2-w_3^2)^2\left(\frac{\xi_1}{w_2w_3}\right)^2\\
      {}+
      \frac{1}{4}(w_3^2-w_1^2)(3w_1^2+w_3^2)\left(\frac{\xi_2}{w_3w_1}\right)^2
\\
      {}+
      \frac{1}{4}(w_2^2-w_1^2)(3w_1^2+w_2^2)\left(\frac{\xi_3}{w_1w_2}\right)^2,
\\
\dot{\alpha}_2=-\alpha_3\alpha_1+\alpha_2(\alpha_3+\alpha_1)+
      \frac{1}{4}(w_3^2-w_1^2)^2\left(\frac{\xi_2}{w_3 w_1}\right)^2\\
      {}+
      \frac{1}{4}(w_1^2-w_2^2)(3w_2^2+w_1^2)\left(\frac{\xi_3}{w_1 w_2}\right)^2\\
      {}+
      \frac{1}{4}(w_3^2-w_2^2)(3w_2^2+w_3^2)\left(\frac{\xi_1}{w_2w_3}\right)^2,\\
\dot{\alpha}_3=-\alpha_1\alpha_2+\alpha_3(\alpha_1+\alpha_2)+
      \frac{1}{4}(w_1^2-w_2^2)^2\left(\frac{\xi_3}{w_1w_2}\right)^2\\
      {}+
      \frac{1}{4}(w_2^2-w_3^2)(3w_3^2+w_2^2)\left(\frac{\xi_1}{w_2w_3}\right)^2\\
      {}+
      \frac{1}{4}(w_1^2-w_3^2)(3w_3^2+w_1^2)\left(\frac{\xi_2}{w_3w_1}\right)^2,
\end{split}\label{sd2}
\end{align}
and
\begin{align}
\begin{split}
(w_2^2-w_3^2)\frac{d}{dt}\left(\frac{\xi_1}{w_2w_3}\right)=&
      \frac{\xi_2}{w_3w_1}\frac{\xi_3}{w_1w_2}(-2w_2^2w_3^2+w_3^2w_1^2+w_1^2w_2^2)\\
      &{}+
      \frac{\xi_1}{w_2w_3}(\alpha_2w_2^2-\alpha_3w_3^2+3\alpha_2w_3^2+3\alpha_3w_2^2),\\
(w_3^2-w_1^2)\frac{d}{dt}\left(\frac{\xi_2}{w_3 w_1}\right)=&
      \frac{\xi_3}{w_1 w_2}\frac{\xi_1}{w_2w_3}(-2w_3^2w_1^2+w_1^2w_2^2+w_2^2w_3^2)\\
      &{}+
      \frac{\xi_2}{w_3w_1}(\alpha_3w_3^2-\alpha_1w_1^2+3\alpha_3w_1^2+3\alpha_1w_2^2),\\
(w_1^2-w_2^2)\frac{d}{dt}\left(\frac{\xi_3}{w_1w_2}\right)=&
      \frac{\xi_1}{w_1w_3}\frac{\xi_2}{w_3w_1}(-2w_1^2w_2^2+w_2^2w_3^2+w_3^2w_1^2)\\
      &{}+
      \frac{\xi_3}{w_1w_2}(\alpha_1w_1^2-\alpha_2w_2^2+3\alpha_1w_2^2+3\alpha_2w_1^2).
\end{split}\label{sd3}
\end{align}

\begin{rem}
If $\xi_1=0,$ $\xi_2=0$ and $\xi_3=0$ then the system of equations \eqref{sd1}, 
\eqref{sd2} and \eqref{sd3} reduces to a sixth-order system \eqref{d-sd}
given by Tod $\cite{Tod}$. Furthermore, if $\alpha_1=w_1, \alpha_2=w_2, \alpha_3=w_3$ 
then \eqref{sd1},\eqref{sd2},\eqref{sd3} reduce to a third-order 
system which determines Atiyah-Hitchin family $\cite{AH}$, and 
if $\alpha_1=0, \alpha_2=0, \alpha_3=0$ then 
the system reduces to a third-order system which determines
BGPP family $\cite{BGPP}$.
\end{rem}

\begin{rem}
If $w_2=w_3$, then we can set $\xi_1=0$, $\xi_2=0$ and $\xi_3=0$ by 
taking another flame. This
is also a diagonal case. Therefore we assume $(w_2-w_3)(w_3-w_1)(w_1-w_2)\neq 0$.
\end{rem}

\section{The Isomonodromic Deformations} \label{Painleve}
Let $(M,g)$ be an oriented Riemannian four manifold. We define a 
manifold $Z$ to be the unit sphere bundle in the bundle of self-dual 
two-forms, and let $\pi: Z\to M$ denote the projection. Each point $z$ 
in the fiber over $\pi(z)$ defines a complex structure on the tangent 
space $T_{\pi(z)}M$, compatible with the metric and its orientation.

Using the Levi-Civita connection, we can split the tangent space $T_z Z$ into 
horizontal and vertical spaces, and the projection $\pi$ identifies the horizontal
space with $T_{\pi(z)}M$. This space has a complex structure defined by $z$ and 
the vertical space is the tangent space of the fiber $S^2\cong \mathbb{CP}^1$ 
which has its natural complex structure. 

The almost complex structure on $Z$ is integrable if and only if the 
metric is anti-self-dual \cite{AHS, Penrose}. In this situation $Z$ is 
called the twistor space of $(M,g)$ and the fibers are called the real 
twistor lines. 

The almost complex structure on $Z$ can be determined by the following
$(1,0)$-forms: 
\begin{align}
\begin{split}
\Theta_1=&z(e^1+\sqrt{-1}e^2)-(e^0+\sqrt{-1}e^3),\\
\Theta_2=&z(e^0-\sqrt{-1}e^3)+(e^1-\sqrt{-1}e^2),\\
\Theta_3=&dz + \frac{1}{2}z^2(\omega^0_3-\omega^1_2+\sqrt{-1}(\omega^0_1-\omega^2_3))\\
	&{}-\sqrt{-1}z(\omega^0_2-\omega^3_1)
	+\frac{1}{2}(\omega^0_1-\omega^1_2-\sqrt{-1}(\omega^0_1-\omega^2_3)),
\end{split}\label{pfaff}
\end{align}
where $\{e^0,e^1,e^2,e^3\}$ is an orthonormal flame, and $\omega^i_j$ are the
connection forms determined by $d e^i+\omega^i_j\wedge e^j=0$ and
$ \omega^i_j+\omega^j_i=0$.
Then the anti-self-dual condition is 
\begin{align}
d\Theta_1&\equiv 0,& d\Theta_2&\equiv 0,& d\Theta_3&\equiv 0 
	&(\textrm{mod}\; \Theta_1, \Theta_2,\Theta_3).
\end{align}

\begin{thm} \label{conj}
If the metric is positive definite, then the Pfaffian is invariant 
under conjugate action and $z\to-{1}/{\bar{z}}$ $\cite{AHS}$. 
If the metric has a signature $(+,+,-,-)$, then the Pfaffian is invariant under 
conjugate action and $z\to\bar{z}$.
\end{thm}

If the metric is $SU(2)$ invariant, we can write
\begin{align}
\left(
  \begin{array}{c}
    \Theta_1   \\
    \Theta_2   \\
    \Theta_3   \\
  \end{array}
\right)=
\left(
  \begin{array}{c}
    0   \\
    0   \\
    1   \\
  \end{array}
\right)dz+
\left(
  \begin{array}{c}
    v_1   \\
    v_2   \\
    v_3   \\
  \end{array}
\right)dt+A\left(
  \begin{array}{c}
    \sigma_1   \\
    \sigma_2   \\
    \sigma_3   \\
  \end{array}
\right),
\end{align}
where $v_1=v_1(z,t)$, $v_2=v_2(z,t)$, $v_3=v_3(z,t)$; $A=\left(a_{i\,j}(z,t)\right)_{i,j=1,2,3}$.

If $\textrm{det}A\equiv 0$, then metric turns to be diagonal, and the 
metric is in the BGPP family \cite{BGPP}. 

If $\textrm{det}A\neq 0$, then we can write
\begin{align}
\left(
  \begin{array}{c}
    \sigma_1   \\
    \sigma_2   \\
    \sigma_3   \\
  \end{array}
\right)&\equiv -A^{-1}\left(
\left(
  \begin{array}{c}
    0  \\
    0  \\
    1  \\
  \end{array}
\right)dz+
\left(
  \begin{array}{c}
    v_1   \\
    v_2   \\
    v_3   \\
  \end{array}
\right)dt
\right), & \mathrm{mod} \; \Theta_1, \Theta_2, \Theta_3.
\end{align}
If we set
\begin{align}
\left(
  \begin{array}{c}
    s_1   \\
    s_2   \\
    s_3   \\
  \end{array}
\right):=
-A^{-1}\left(
\left(
  \begin{array}{c}
    0  \\
    0  \\
    1  \\
  \end{array}
\right)dz+
\left(
  \begin{array}{c}
    v_1   \\
    v_2   \\
    v_3   \\
  \end{array}
\right)dt
\right),
\end{align}
then 
\begin{align}
d\left(
  \begin{array}{c}
    s_1   \\
    s_2   \\
    s_3   \\
  \end{array}
\right)&\equiv \left(
  \begin{array}{c}
    s_2\wedge s_3   \\
    s_3\wedge s_1   \\
    s_1\wedge s_2   \\
  \end{array}
\right),& \mathrm{mod} \; \Theta_1, \Theta_2, \Theta_3. \label{sigmahat}
\end{align}
Since $s_1,s_2,s_3$ are one-forms on 
$(z,t)-$plane, the congruency equation \eqref{sigmahat} turns to be a plain equation:
\begin{align}
d\left(
  \begin{array}{c}
 s_1      \\
 s_2      \\
 s_3      \\
  \end{array}
\right)=\left(
  \begin{array}{c}
    s_2\wedge s_3   \\
    s_3\wedge s_1   \\
    s_1\wedge s_2   \\
  \end{array}
\right).
\end{align}
 If the metric is positive definite, then 
 $s_1,s_2,s_3$ are invariant under conjugate action and 
 $z\to-1/\bar{z}$ by theorem \ref{conj}.
 And if the metric has a signature $(+,+,-,-)$, then 
 $s_1,s_2,s_3$ are invariant under conjugate action and $z\to\bar{z}$ . 

If we set
\begin{align}
\Sigma&=\frac{1}{\sqrt{2}}\left(
  \begin{array}{cc}
    \sqrt{-1}s_1   &-s_3+\sqrt{-1}s_{2}    \\
    s_3+\sqrt{-1}s_2   &-\sqrt{-1}s_1    \\
  \end{array}
\right)\\
&=:{}-B_1\,dz-B_2\,dt,
\end{align}
then
\begin{align}
d\Sigma+\Sigma\wedge\Sigma=0.
\end{align}
This is the isomonodromy condition for the the following linear problem \cite{JMU-I}
\begin{align}
\left(\frac{d}{dz}-B_1\right)\left(
  \begin{array}{c}
    y_1   \\
    y_2   \\
  \end{array}
\right)=0. \label{isom}
\end{align}

\begin{lemma}\label{detAlemma}
The components of $B_1$ are rational functions of $z$,
\[
 B_1=\frac{F(z)}{G(z)},
\]
where $F(z)$ is degree $2$ and $G(z)$ is degree $4$. If the metric is 
positive definite, then $B_1\to-{}^tB_1$ under conjugate action and 
 $z\to-1/\bar{z}$. And if the metric has a signature $(+,+,-,-)$, then 
 $B_1\to-{}^tB_1$  under conjugate action and  $z\to\bar{z}$.
\end{lemma}
For this lemma, generically $B_1$ has four simple poles. In this case, the 
deformation equation of \eqref{isom} is Painlev\'e VI.
\begin{thm}
The anti-self-dual equations on $SU(2)$-invariant metrics generically reduce to
Painlev\'e VI.
\end{thm}

The idea of Hitchin $\cite{Hitchin}$ is that the lifted action of 
$SU(2)$ on the twistor space $Z$ gives a homomorphism of vector bundles 
$\alpha: Z \times\mathfrak{su}(2)^\mathbb{C}\to TZ$, and the inverse of 
$\alpha$ gives a flat meromorphic $SL(2,\mathbb{C})$-connection, which 
determine isomonodromic deformations. Since one-forms 
$\Theta_1,\Theta_2,\Theta_3$ on $Z$ can be considered as are 
infinitesimal variations, we can identify $\Sigma$ with $\alpha^{-1}$.

First, we review the positive definite metric.
By lemma \ref{detAlemma} the poles of $B_1$ make antipodal pairs 
$\zeta_0, -1/\bar{\zeta_0}$, and $\zeta_1, -1/\bar{\zeta_1}$ on 
$\mathbb{CP}^1$. 

Therefore, we have two types of configuration of poles of $B_1$:
\begin{enumerate}
\item[(a)] $B_1$ has four simple poles $\zeta_0,$ $\bar{\zeta_0},$ $\zeta_1,$ 
$\bar{\zeta_1}$ on $\mathbb{C}\setminus\mathbb{R}$. 
\[
 B_1=\frac{A_0}{z-\zeta_0} + \frac{-{}^t\bar{A_0}}{z+1/\bar{\zeta_0}}+
 \frac{A_1}{z-\zeta_1} + \frac{-{}^t\bar{A_1}}{z+1/\bar{\zeta_1}}.
\] 
The deformation equation is Painlev\'e VI with parameters,
 \[
  \left(\alpha,\beta,\gamma,\delta\right)=
  \left(\frac{1}{2}(\theta_0-1)^2, \frac{1}{2}\bar{\theta_0}^2, 
  -\frac{1}{2}\theta_1^2,\frac{1}{2}(1+\bar{\theta_1}^2)\right),
 \] 
where $\theta_0^2=2\,\mathrm{tr}A_0^2$, $\theta_1^2=2\,\mathrm{tr}A_1^2$.
\item[(b)] $B_1$ has two double poles $\zeta,$ $\bar{\zeta}$ on $\mathbb{C}\setminus\mathbb{R}$.
\[
 B_1=\frac{A_2}{(z-\zeta)^2}+\frac{\sqrt{-1}C}{z-\zeta}+
 \frac{-\sqrt{-1}C}{z+1/\bar{\zeta}}+
 \frac{-{}^t\bar{A_2}/\bar{\zeta}^2}{(z+1/\bar{\zeta})^2},
\]
where $C=-{}^t\bar{C}$. The deformation equation is Painlev\'e III 
with parameters,
\[
 \left(\alpha,\beta,\gamma,\delta\right)=
 \left(4\theta,4(1+\bar{\theta}),4,-4\right),
\]
where $\theta^2={2(\mathrm{tr}(A_3C))^2}/{\mathrm{tr}C^2}$.
\end{enumerate}

\begin{thm} \label{diagonal-case}
 If the metric is positive definite, the anti-self-dual equations reduce 
 to the following two Painlev\'e equations:
\begin{enumerate}
\item[$(\mathrm{a})$]A family of Painlev\'e VI with two complex parameters,
 \[
  \left(\alpha,\beta,\gamma,\delta\right)=
  \left(\frac{1}{2}(\theta_0-1)^2, \frac{1}{2}\bar{\theta_0}^2, 
  -\frac{1}{2}\theta_1^2,\frac{1}{2}(1+\bar{\theta_1}^2)\right),
 \] 
\item[$(\mathrm{b})$]A family of Painlev\'e III with one complex parameter,
\[
 \left(\alpha,\beta,\gamma,\delta\right)=
 \left(4\theta,4(1+\bar{\theta}),4,-4\right).
\]
\end{enumerate}
\end{thm}

\begin{rem}
 Hitchin shows the anti-self-dual equations reduce to Painlev\'e VI with 
 the parameters above \ $\mathrm{(\cite{Hitchin}, P.50\ (14))}$. 
 Dancer $\cite{Dancer}$ shows the scalar-flat diagonal K\"ahler metric 
 is specified by a solution of 
 Painlev\'e III with parameters $(\alpha,\beta,\gamma,\delta)=(4,4,4,-4)$.
 Now, theorem $\ref{diagonal-case}\ \mathrm{(b)}$ is a generalization of 
 Dancer's result. 
\end{rem}

From now on, we will classify the anti-self-dual metrics 
with a signature $(+,$ $+,$ $-,$ $-)$. 
By lemma \ref{detAlemma} the poles of $B_1$ make conjugate pairs 
$\zeta_0, \bar{\zeta_0}$, and $\zeta_1, \bar{\zeta_1}$ in 
$\mathbb{CP}^1$. Therefore we have five types of metrics corresponding 
to configuration of poles of $B_1$.

\begin{thm} \label{5cases}
If the metric has a signature $(+,+,-,-)$, the anti-self-dual equations 
reduce to the following five Painlev\'e equations:
\begin{enumerate}
\item[$(\mathrm{a})$]
 Painlev\'e VI with two complex parameters 
 \[
  \left(\alpha,\beta,\gamma,\delta\right)=
  \left(\frac{1}{2}(\theta_0-1)^2, \frac{1}{2}\bar{\theta_0}^2, 
  -\frac{1}{2}\theta_1^2,\frac{1}{2}(1+\bar{\theta_1}^2)\right),
 \] 
\item[$(\mathrm{b})$]
 Painlev\'e V with one real and one complex parameters 
\[
  \left(\alpha,\beta,\gamma,\delta\right)= 
  \left(\frac{1}{2}(\theta_0+\bar{\theta_0}+\theta_\infty)^2,
 -\frac{1}{2}(\theta_0+\bar{\theta_0}-\theta_\infty)^2, 
 1-\theta_0+\bar{\theta_0},\frac{1}{2}
  \right),
\]
where $\theta_\infty\in\mathbb{R}$.
\item[$(\mathrm{c})$]
 Painlev\'e III with one complex parameter 
\[
 \left(\alpha,\beta,\gamma,\delta\right)=
 \left(4\theta,4(1+\bar{\theta}),4,-4\right).
\]
\item[$(\mathrm{d})$]
 Painlev\'e III with with two real parameters 
\[
 \left(\alpha,\beta,\gamma,\delta\right)=
 \left(4\theta_1,4(1+{\theta_2}),4,-4\right).
\]
\item[$(\mathrm{e})$]
Painlev\'e II with one real parameter $\alpha$.
\end{enumerate}
\end{thm}

Proof.

Since the poles of $B_1$ make conjugate pairs 
$\zeta_0, \bar{\zeta_0}$ and $\zeta_1, \bar{\zeta_1}$, we have 
five types of configuration of poles of $B_1$. In each case, we can 
calculate local exponents at singularities. These local exponents 
corresponding to parameters of Painlev\'e equations (see \cite{JM-II}).
\begin{enumerate}
\item[(a)] Generically, 
 $B_1$ has four simple poles $\zeta_0,$ $\bar{\zeta_0},$ $\zeta_1,$ 
$\bar{\zeta_1}$ on $\mathbb{C}\setminus\mathbb{R}$. 
\[
 B_1=\frac{A_0}{z-\zeta_0} + \frac{-{}^t\bar{A_0}}{z-\bar{\zeta_0}}+
\frac{A_1}{z-\zeta_1} + \frac{-{}^t\bar{A_1}}{z-\bar{\zeta_1}}.
\]
The deformation equations are Painlev\'e VI  with parameters,
  \[
  \left(\alpha,\beta,\gamma,\delta\right)=
  \left(\frac{1}{2}(\theta_0-1)^2, \frac{1}{2}\bar{\theta_0}^2, 
  -\frac{1}{2}\theta_1^2,\frac{1}{2}(1+\bar{\theta_1}^2)\right),
 \]  
where $\theta_0^2=2\,\mathrm{tr}A_0^2$, $\theta_1^2=2\,\mathrm{tr}A_1^2$.
\item[(b)] If $\zeta_0=\bar{\zeta_0}(=\eta)$, 
then $B_1$ has two simple poles $\zeta_1,$ $\bar{\zeta_1}$ on 
$\mathbb{C}\setminus\mathbb{R}$, and one double pole $\eta$ on $\mathbb{R}$.  
\[
 B_1=\frac{C}{{\left( z - \eta  \right) }^2} + 
  \frac{-A_2 +{}^t\bar{A_2}}{z - \eta } + \frac{A_2}{z - \zeta_1 } + 
  \frac{-{}^t\bar{A_2}}{z - {\bar{\zeta_1 }}},
\]
where $C=-{}^t\bar{C}$. 
The deformation equation is Painlev\'e V with parameters,
\[
  \left(\alpha,\beta,\gamma,\delta\right)= 
  \left(\frac{1}{2}(\theta_0+\bar{\theta_0}+\theta_\infty)^2,
 -\frac{1}{2}(\theta_0+\bar{\theta_0}-\theta_\infty)^2, 
 1-\theta_0+\bar{\theta_0},\frac{1}{2}
  \right),
\]
where $\theta_0^2=2\,\mathrm{tr}A_2^2$, 
$\theta_\infty^2
=2\left(\mathrm{tr}\left(A_2-{}^t\bar{A_2}\right)C\right)^2/\mathrm{tr}{C^2}$.
\item[(c)] If $\zeta_0=\zeta_1(=\zeta)$, then 
$B_1$ has two double poles $\zeta,$ $\bar{\zeta}$ on 
$\mathbb{C}\setminus\mathbb{R}$.
\[
 B_1=\frac{A_3}{(z-\zeta)^2}+\frac{\sqrt{-1}C}{z-\zeta}+\frac{-\sqrt{-
 1}C}{z-\bar{\zeta}}+\frac{-{}^t\bar{A_3}}{(z-\bar{\zeta})^2}, 
\] 
where $C=-{}^t\bar{C}$. The deformation equation is 
Painlev\'e III with parameters,
\[
 \left(\alpha,\beta,\gamma,\delta\right)=
 \left(4\theta,4(1+\bar{\theta}),4,-4\right),
\]
where $\theta^2={2(\mathrm{tr}A_3C)^2}/{\mathrm{tr}C^2}$.
\item[(d)] If $\zeta_0=\bar{\zeta_0}(=\eta_0)$, 
$\zeta_1=\bar{\zeta_1}(=\eta_1)$, then
 $B_1$ has two double poles $\eta_0,$ $\eta_1$ on $\mathbb{R}$.
\[
 B_1=\frac{C_1}{(z-\eta_0)^2}+\frac{C_2}{z-\eta_0}+
  \frac{-C_2}{z-\eta_1}+\frac{C_3}{(z-\eta_1)^2}, 
\] 
where $C_i=-{}^t\bar{C_i}$ $(i=1,2,3)$. The deformation equation is 
Painlev\'e III with parameters,
\[
 \left(\alpha,\beta,\gamma,\delta\right)=
 \left(4\theta_1,4(1+{\theta_2}),4,-4\right),
\]
where $\theta_1^2={2(\mathrm{tr}C_1C_2)^2}/{\mathrm{tr}C_2^2}$,
$\theta_2^2={2(\mathrm{tr}C_2C_3)^2}/{\mathrm{tr}C_2^2}$.
\item[(e)] If $\zeta_0=\bar{\zeta_0}=\zeta_1=\bar{\zeta_1}(=\eta)$, then 
$B_1$ has one quadruple pole $\eta$ on $\mathbb{R}$.
\[
 B_1=\frac{C_1}{{\left( z - \eta  \right) }^4} + 
  \frac{C_2}{{\left( z - \eta \right) }^3} + 
  \frac{C_3}{{\left( z - \eta \right) }^2},
\]
where $C_i=-{}^t\bar{C_i}$ $(i=1,2,3)$. If $\mathrm{det}C_1\neq 0$, then the 
deformation equation is Painlev\'e II with a parameter,
\[
 \alpha = \frac{1}{2}(1+\mathrm{tr}C_2C_3).
\] If $\mathrm{det}C_1=0$, then the 
deformation equation is Painlev\'e I, but since $C_1=-{}^t\bar{C_1}$, 
this never occurs. \hfill\rule{3pt}{8pt}
\end{enumerate}

\section{Hermitian Structure and Null Surfaces}
In this section we will consider the geometric meaning of the metrics 
corresponding with each type of Painlev\'e functions.

\begin{lemma}\label{multi-p}
 Let $g$ be a anti-self-dual metric. 
 If $B_1$ has a multiple pole, then the 
 Pfaffian $\Theta_1|_{z=\zeta(t)},$ 
 $\Theta_2|_{z=\zeta(t)}$ is integrable. 
 Conversely, if the Pfaffian 
 $\Theta_1|_{z=\zeta(t)},$ $\Theta_2|_{z=\zeta(t)}$ is integrable for 
 some $z=\zeta(t)$, then $B_1$ has a multiple pole on $z=\zeta(t)$.
\end{lemma}

Proof.

If $z=\zeta(t)$ is a multiple zero of $G(z)$, then $G|_{z=\zeta(t)}$ and 
$dG|_{z=\zeta(t)}$ must be vanish. Furthermore, $\Theta_3\equiv 0$ $(\textrm{mod}$ 
$\Theta_1,$ $\Theta_2,$ $G,$ $dG)$. Therefore, the Pfaffian 
${\Theta_1|_{z=\zeta(t)},\Theta_2|_{z=\zeta(t)}}$ is integrable. 

Conversely,
if the Pfaffian $\Theta_1|_{z=\zeta(t)},$ $\Theta_2|_{z=\zeta(t)}$ is integrable for 
some $z=\zeta(t)$, then $\Theta_3|_{z=\zeta(t)}\equiv 0$ $(\mathrm{mod} 
\Theta_1,\Theta_2)$. Therefore the denominator $G$ of $B_1$ has zero on 
$z=\zeta(t)$. Furthermore, $\Theta_3|_{z=\zeta(t)}\equiv 0$ $(\mathrm{mod} 
\Theta_1,\Theta_2, G)$ is equivarent to $dG|_{z=\zeta(t)}= 0$, 
therefore $z=\zeta(t)$ is a double pole of $B_1$. 
\hfill\rule{3pt}{8pt}

~

In \cite{oku} we show that if a positive definite $SU(2)$ invariant 
anti-self-dual metric is corresponding with Painlev\'e III with one complex 
parameter $(4\theta,$ $4(1+\bar{\theta}),$ $4,$ $-4)$, then the Pfaffian 
${\Theta_1|_{z=\zeta(t)},\Theta_2|_{z=\zeta(t)}}$ ($z=\zeta(t)$ is 
a double pole of $B_1$) determines a $SU(2)$ invariant hermitian 
structure. 

In the same way, we have the following theorem for the 
metric with a signature $(+,+,-,-)$ .

\begin{thm}
 If  the anti-self-dual equations reduce to Painlev\'e III with one complex 
 parameter $(4\theta,$ $4(1+\bar{\theta}),$ $4,$ $-4)$, then there 
 exists a $SU(2)$-invariant hermitian structure. Conversely, if there 
 exists a  $SU(2)$-invariant hermitian structure, then the 
 anti-self-dual equations reduce to Painlev\'e III with parameters above.
\end{thm}

\begin{definition}
If $g(X,X)=0$, then $X\in TM$ is said to be a null direction.
\end{definition}

The Pfaffian 
${\Theta_1|_{z=\eta(t)},\Theta_2|_{z=\eta(t)}}(\eta(t)\in\mathbb{R})$ 
determines two dimensional null directions on $TM$.

\begin{definition}
 Let $N$ be a two dimensional subspace of $M$. If $g(X,X)=0$ 
 for any $X\in  TN$, then N is called a null surface.
\end{definition}

From lemma \ref{multi-p}, if $z=\eta(t)\in\mathbb{R}$ is a multiple pole 
of $B_1$, then the Pfaffian ${\Theta_1|_{z=\eta(t)},\Theta_2|_{z=\eta(t)}}$ is 
integrable, and then for any  $x\in M$ there 
exists a $SU(2)$-invariant null surface passing through $x$. 
Conversely, 
for any $x\in M$, if there exists a $SU(2)$-invariant null surface passing 
through $x$, then the null surface is represented by 
Pfaffian ${\Theta_1|_{z=\eta(t)},\Theta_2|_{z=\eta(t)}}$ for some 
$\eta(t)\in\mathbb{R}$, and then $z=\eta(t)$ is a multiple pole 
of $B_1$

\begin{thm}
 If the anti-self-dual equations reduce to Painlev\'e V with one real 
 and one complex parameters 
$(\frac{1}{2}(\theta_0+\bar{\theta_0}+\theta_\infty)^2,$
$-\frac{1}{2}(\theta_0+\bar{\theta_0}-\theta_\infty)^2,$
$1-\theta_0+\bar{\theta_0},$ $\frac{1}{2})$, where $\theta_\infty \in \mathbb{R}$,
or Painlev\'e II with one real parameter $\alpha$,
then for any $x\in M$ there exists one $SU(2)$-invariant null surface 
passing through $x$. Conversely, for any $x\in M$ there exists one 
$SU(2)$-invariant null surface passing through $x$, then the 
anti-self-dual reduce to Painlev\'e V or II with parameters above.

If the anti-self-dual equations reduce to Painlev\'e III with two real 
parameters $(4\theta_1,4(1-\theta_2),4,-4)$, then for any $x\in M$ there 
exist two $SU(2)$-invariant null surfaces passing through $x$. 
Conversely, for any $x\in M$ there 
exist two $SU(2)$-invariant null surfaces passing through $x$, then the 
anti-self-dual reduce to Painlev\'e III with parameters above.

\end{thm}

\section{Summary}

We classified the $SU(2)$-invariant anti-self-dual metric with a 
signature $(+,+,-,-)$ into the five cases (a)--(e) (theorem \ref{5cases}). 
The meaning of the types of Painlev\'e equations are as follows: 
\begin{enumerate}
\item
Generically, the anti-self-dual metric are specified by a solution of 
Painlev\'e VI with two complex parameters. 
\item
If the anti-self-dual metric is 
specified by a solution of Painlev\'e III with two complex parameters, 
then there exist a $SU(2)$-invariant hermitian structure. 
\item
If the anti-self-dual metric is 
specified by a solution of Painlev\'e V with one real and two complex 
parameters, or Painlev\'e II with one real parameter, then for any $x\in 
M$ there exists one real null surface passing through $x$.
\item
If the anti-self-dual metric is 
specified by a solution of Painlev\'e III with two real parameters, 
then for any $x\in M$ there exist two real null surfaces passing through $x$.
\end{enumerate}

\section{Appendix}

We review the Painlev\'e equations, second order nonlinear differential 
equations without moving critical points. We list six equations 
classified by Painlev\'e and Gambier, where $\alpha,\beta,\gamma,\delta$ 
are parameters \cite{Painleve}. 

\begin{enumerate}
\item Painlev\'e I 
\[
 \frac{d^2 q}{d x^2}=6q^2+x.
\]
\item Painlev\'e II
\[
 \frac{d^2 q}{dx^2}=2 q^3+x q + \alpha.
\]
\item Painlev\'e III
\[
 \frac{d^2q}{dx^2}=\frac{1}{q}\left(\frac{dq}{dx}\right)^2
 -\frac{1}{x}\frac{dq}{dt}+\frac{1}{x}\left(\alpha q^2+\beta\right)
 +\gamma q^3 +\frac{\delta}{q}.
\]
\item Painlev\'e IV
 \[
	\frac{d^2q}{dx^2}=\frac{1}{2q}\left(\frac{dq}{dx}\right)^2
     +\frac{3}{2}q^3+4xq^2+2(x^2-\alpha)q+\frac{\beta}{q}.
 \]
\item Painlev\'e V
\begin{multline*}
	\frac{d^2 q}{d x^2} = 
  \left(\frac{1}{2 q}+\frac{1}{q-1}\right)\left(\frac{dq}{dx}\right)^2
   -\frac{1}{x}\frac{dq}{dx} +
   \frac{(q-1)^2}{x^2}
   \left(\alpha q + \frac{\beta}{q}\right)+
   \frac{\gamma q}{x}+\frac{\delta q(q+1)}{q-1}.
\end{multline*}
\item Painlev\'e VI
 \begin{multline*}
   \frac{d^2 q}{d x^2} = \frac{1}{2}\left( \frac{1}{q} +
       \frac{1}{q-1} + \frac{1}{q-x} \right) \left(\frac{d q }{d x}\right)^2
            \! - \left( \frac{1}{x}+\frac{1}{x-1}
               +\frac{1}{q-x} \right)\frac{d q }{d x}  \\
      + \frac{q\left(q-1\right)\left(q-x\right)}{x^2\left(x-1\right)^2}
        \left\{  \alpha + \beta\frac{x}{q^2} +
            \gamma\frac{x-1}{\left(q-1\right)^2 }+
             \delta \frac{x\left(x-1\right)}{\left(q-x\right)^2} \right\}.
 \end{multline*}
\end{enumerate}
%%%%%%%%%%%%%%%%%%%%%%%%%%%%%%%%%%%%%%%%%%%%%%%%%%%%%%%%%%
%%%%%%%%%%%%%%%%%%%%%%%%%%%%%%%%%%%%%%%%%%%%%%%%%%%%%%%%%%
%%%%%%%%%%%%%%%%%%%%%%%%%%%%%%%%%%%%%%%%%%%%%%%%%%%%%%%%%%


\begin{thebibliography}{99}
\bibitem{AH}Atiyah, M. F. and Hitchin, N. J.: 
	Low energy scattering of non-Abelian monopoles, 
	\textit{Phys. Lett. A} \textbf{107} (1985), 21--25
\bibitem{AHS}Atiyha, M. F., Hitchin, N. J. and Singer, I. M.: 
	Self-duality in four-dimensional Riemannian geometry, 
	\textit{Proc. Roy. Soc, London Ser. A} \textbf{362} (1978) 425--461
%\bibitem{Besse}Besse, A. L.: Einstein Manifolds, Springer (1987)
\bibitem{BGPP}Belinski, V. A., Gibbons, G. W., Page, D. W., Page, D. W. and Pope, C. N.: 
	Asymptotically Euclidean Bianchi IX metrics in quantum gravity, 
	\textit{Phys. Lett. B} \textbf{76} (1978),433--435
\bibitem{Dancer}Dancer, A. S.: 
	Scalar-flat K\"ahler metrics with $SU(2)$ symmetry, 
	\textit{J.reine. angew. Math.} \textbf{479} (1996), 99--120
\bibitem{JMU-I}Jimbo, M., Miwa, T. and Ueno, K.:
	Monodromy Preserving Deformation of Linear Ordinary Differential 
	Equations with Rational Coefficients. I,
	\textit{Physica} \textbf{2D} (1981), 306--352
\bibitem{JM-II}Jimbo, M. and Miwa, T.:
	Monodromy Preserving Deformation of Linear Ordinary Differential 
	Equations with Rational Coefficients. II,
	\textit{Physica} \textbf{2D} (1981), 407--448
\bibitem{Hitchin}Hitchin, N. J.: 
	Twistor spaces, Einstein metrics and isomonodromic deformations, 
	\textit{J. Differential Geom.} \textbf{42} (1995), 30--112
\bibitem{oku}Okumura, S.:
	The Self-Dual Hermitian Metric and Painlev\'e III, preprint
\bibitem{Painleve}Painlev\'e, P.:
	Sur les \'equations diff\'erentielles du second ordre \`a points 
	critiques fixes, 
	\textit{C. R. Acad. Sci. Paris} \textbf{143} (1906), 1111--1117
%\bibitem{PP}Pedersen, H. and Poon, Y. S.: 
%	K\"ahler surfaces with zero scalar curvature, 
%	\textit{Classical Quantum Gravity} \textbf{7} (1990), 1707--1719
\bibitem{Penrose}Penrose, R.: 
	Nonlinear gravitons and curved twistor theory, 
	\textit{Gen. Rel. Grav.} \textbf{7} (1976), 31--52
\bibitem{Tod}Tod, K. P.: 
	Self-dual Einstein metrics from the Painlev\'e VI equation, 
	\textit{Phys. Lett. A} \textbf{190} (1994), 221--224

\end{thebibliography}
\end{document}